\documentclass[journal, onecolumn, 12pt]{IEEEtran}

\usepackage{amsmath,amssymb,cite}
\usepackage[dvips]{graphicx}
\usepackage[usenames]{color}
\usepackage{amsfonts}
\usepackage{latexsym}
\usepackage{multirow}
\usepackage{subfigure}
\usepackage{float}
\newtheorem{lemma}{{\textit{{Lemma}}}}

\newtheorem{remark}{\textit{{Remark}}}
\newtheorem{proof}{\textit{{Proof}}}

\newtheorem{example}{\textit{{Example}}}

\begin{document}

\title{Low-PAPR Preamble Design for FBMC Systems}

\author{Zilong~Liu,~Pei~Xiao,~Su~Hu
\thanks{Zilong Liu and Pei Xiao are with Institute of Communication Systems, 5G Innovation Centre, University of Surrey, UK (E-mail: {zilong.liu@surrey.ac.uk, p.xiao@surrey.ac.uk}). Su Hu is with National Key Laboratory on Communications, University of Electronic Science and Technology of China, China. Email: husu@uestc.edu.cn.}
\thanks{The work of Z. Liu and P. Xiao was supported in part by the EPSRC project ``New Air Interface Techniques for Future Massive Machine Communications" (EP/P03456X/1) and the H2020 EU-Taiwan project ``Converged Wireless Access for Reliable 5G MTC for Factories of Future - Clear5G" (61745). The work of S. Hu was supported in part by the Ministry of Science and Technology of China
(MOST) Program of International S\&T Cooperation under Grant 2016YFE0123200. The work of Z. Liu and S. Hu was also supported in part by National Natural Science Foundation of China through the Research Fund for International Young Scientists under Grant 61750110527.}
}

\maketitle

\begin{abstract}
This paper presents a family of training preambles for offset QAM (OQAM) based filter-bank multi-carrier (FBMC) modulations with low peak-to-average power ratio (PAPR) property. We propose to use binary Golay sequences as FBMC preambles and analyze the maximum PAPR for different numbers of zero guard symbols. For both the PHYDYAS and Hermite prototype filters with overlapping factor of 4, as an illustration of the proposed preambles, we show that a preamble PAPR less than 3 dB can be achieved with probability of one, when three or more zero guard symbols are inserted in the vicinity of each preamble.
\end{abstract}


\section{Introduction}
FBMC with OQAM is a multicarrier modulation scheme which relies on pulses with good time-frequency localization property \cite{Farhang2011}. From hereon, we shall use FBMC to denote ``FBMC/OQAM", although there exist other FBMC variants too. Compared to conventional orthogonal frequency division multiplexing (OFDM) employing cyclic prefix (CP) and rectangular pulses, FBMC possesses higher spectral efficiency as it can work without the need of a CP even with a frequency-selective channel. Moreover, it results in much lower out-of-band power emission owing to the tight spectrum containment of FBMC pulses as well as relatively strong resilience to carrier frequency offsets and Doppler spreads \cite{FBMC-book}. These advantages give FBMC a great potential for the support of a diverse range of modern use cases where flexible time-frequency allocations are highly demanded. While recent advancement of 3GPP indicates that (windowed or filtered) OFDM will be used in 5G [largely for the backwards compatibility with legacy Long Term Evolution (LTE) systems], the voice for adopting FBMC in future mobile networks does not seem to decline \cite{NSR2017}.

A primary feature that distinguishes FBMC from conventional OFDM is that orthogonality among different subcarrier waveforms holds in the real field only. This results in the intrinsic imaginary interference which should be carefully dealt with especially at the channel estimation stage, where channel coefficients need to be estimated in the complex domain. Extensive research efforts have been made to mitigate or overcome this problem in FBMC systems, including both preamble- and scattered pilot-based approaches \cite{FBMC-book}. This paper is concerned with the preamble-based channel estimation approach. Well-known examples include pairs of pilots (POP) \cite{Lele2008}, interference approximation method (IAM) \cite{Lele2008,Lele2008-v2,Du2008}, and interference cancellation method (ICM) \cite{Hu2010}. Kofidis \emph{et al}. present a comprehensive survey in \cite{Kofidis2013} of this topic, including preamble design and associated estimation methods for FBMC systems equipped with multiple antennas. The preamble PAPRs of these methods, however, are mostly overlooked in the design procedure. Recently, two types of low-PAPR preambles have been reported in \cite{Taheri2015} and \cite{Hu2017}. Nevertheless, no analytical evidence has been provided, only numerical results. When a nonlinear power amplifier model is considered, \cite{Levanen2016} points out that the IAM-R and IAM-C preambles\footnote{IAM-R and IAM-C are two types of FBMC preambles with no sparsity in the frequency domain, i.e., all the preamble entries take nonzero values. Examples of IAM-R and IAM-C preambles can be found in Fig. 3 of \cite{Kofidis2013}.} suffer from very poor error rate performance due to their high PAPRs. It should be noted that low-PAPR FBMC preambles are crucial in ensuring a practical system performance. A high-PAPR preamble tends to drive the transmit power amplifier to its nonlinear region, which in turn results in a distorted preamble signal (and hence inaccurate estimates of channel coefficients). To avoid this, certain power back-off may be applied but this could considerably reduce the effective preamble power (again leading to channel estimation performance deterioration).

In this paper, we study the PAPRs of FMBC preambles using the well-known PHYDYAS \cite{PHYDYAS} and Hermite \cite{Hermite} prototype filters. Although there have been substantial research attempts in reducing the PAPRs of the FBMC data payload parts \cite{Jiang2013,Zhao2017,Na2018}, we target at a lower PAPR for FBMC preambles by taking advantage of some coding techniques relevant to Golay sequences \cite{Golay1961}. Found by Marcel Golay in 1951 \cite{Golay-51} in his design of infrared multislit spectrometry, Golay sequences have found diverse applications in wireless communications \cite{Parker02}. A few selected applications in wireless communications include: optimal channel estimation in frequency-selective channels \cite{SpasojevicGeorghiades2001,SQWang2007}, Doppler-resilient waveform design \cite{Budisin91,AAWS2008}, interference-free multi-carrier code-division multiple-access (MC-CDMA) communications \cite{HHC01,HHC-book,Liu-TWC2014}. In particular, in conventional OFDM systems\footnote{From hereon, without any specific announcement, we use ``OFDM" to denote ``conventional OFDM".} with rectangular prototype filter, it is known that each Golay sequence has a PAPR of at most 2 (i.e., 3 dB) if it is spread over the frequency domain \cite{Boyd86,Popovic1991}. In 1999, Davis and Jedwab proposed an explicit, non-recursive construction for Golay sequences, from certain second-order cosets of the first order Reed-Muller code, and then applied them in code-keying OFDM systems for low PAPR transmission \cite{Davis-Jedwab1999}. Despite a rich body of literature on OFDM PAPR reduction using Golay sequences \cite{Paterson00,Tarokh01,Tarokh03,Liu-TCOM2013,Li-GCS-2013,Liu-GCS-2013}, little research has been done on their extension to FBMC systems, to the best of our knowledge. An OFDM system adopts a rectangular prototype filter whose time-domain waveform lasts one symbol only. This makes the PAPR analysis relatively easy as the waveform of every OFDM symbol (preamble or data) is strictly localized within its own time window and with no interference from any neighbouring OFDM symbols. In contrast, the analysis of the preamble PAPR of FBMC systems needs to take into account of the prototype filtering effect and the interference from the data payload part.

%whilst an FBMC system employs a prototype filter   %These complementary sequences are generated via the generalized Boolean functions, and are called Golay-Davis-Jedwab (GDJ) sequences in this paper. They have been applied to the power control in orthogonal frequency-division multiplexing (OFDM) system. The construction of complementary codes, which is a generalization of that in \cite{DAVIS-JEWAB99}, was proposed by Paterson by relating each code with a graph \cite{}. However, Paterson's construction cannot generate a mutually orthogonal complementary code set. This gap was filled by A. Rathinakumar and A. K. Chaturvedi by their CCC construction in \cite{RATHINAKUMAR08}. In addition, there also have been a lot of efforts on the design of quadrature amplitude modulation (QAM)  complementary sequences for OFDM system with higher-order modulations \cite{Tarokh00}-\cite{Huang10}.

We first show in Subsection III-A that the instantaneous magnitudes of the FBMC preamble signals, influenced by data payload in the vicinity of each preamble, can be characterized by Rician distributions. This enables us to calculate the probability of the preamble instantaneous-to-average power ratio (IAPR), whose analytical expression implies that a low preamble PAPR in FBMC is possible provided that a special linear phase transform of such a preamble gives rise to a low PAPR in conventional OFDM systems (see \textit{Remark 2}). Motivated by this interesting observation, we examine the application of Golay sequences for FBMC preamble PAPR reduction. It is shown that FBMC preambles built with binary Golay sequences result in PAPRs that are upper bounded by 1.635 dB and 2.673 dB, for the PHYDYAS and the Hermite prototype filters (both having overlapping factor of 4), respectively, when three or \textit{more} zero guard symbols are placed in the preamble neighborhood. When two zero guard symbols are inserted, our numerical simulations show that less than 3 dB preamble PAPR can be achieved for the PHYDYAS prototype filter with probability higher than $99.99\%$ and for the Hermite prototype filter with probability 1. Another interesting observation made in this paper is that, with a Golay sequence adopted as an FBMC preamble, a prototype filter with better frequency-localization will give rise to a lower PAPR. This can be verified through the PAPR comparison of the PHYDYAS and the Hermite prototype filters.

\section{System Model and Problem Formulation}

\subsection{FBMC System Model}

We consider an FBMC baseband model with $M$ subcarriers, where the subcarrier spacing is $1/T$ with $T$ being the complex symbol interval. The equivalent FBMC signal is expressed as %\cite{Boelcskei}
\begin{equation}\label{TxSIG}
s(t) = \sum\limits_{m = 0}^{M - 1} \sum\limits_{n \in \mathbb{Z}} a_{m,n} g_{m,n} (t),
\end{equation}
where $j=\sqrt{-1}$,
\begin{displaymath}
g_{m,n} (t)=j^{m + n}e^{j2\pi m t/T} g\left(t - nT/2 \right),
\end{displaymath}
$a_{m,n}$ is the real-valued offset-QAM symbol transmitted over the $m$th subcarrier and the $n$th time-slot, and $T/2$ is the interval of real-valued symbols. Moreover, $ g\left (t \right )$ is the employed symmetrical real-valued prototype filter impulse response with total energy of one, and $g_{m, n} \left( t \right)$ represents the synthesis basis which is obtained by the time-frequency translated version of $ g\left(t \right)$, where the transmultiplexer response of $g_{m, n} \left ( t \right )$ is defined as $\zeta^{p,q}_{m,n}=\int g_{m,n}(t)g^*_{p,q}(t)dt$. Note that all the values of $\zeta^{p,q}_{m,n}$ are purely imaginary, except at $m=n, p=q$, implying that FBMC systems enjoy a real-field orthogonality only. Hence, they suffer from \emph{intrinsic imaginary interference} for any $(m,n) \neq (p,q)$, even with a distortion-free channel \cite{FBMC-book}.

We consider the PHYDYAS and Hermite prototype filters, both having identical overlapping factor of $K=4$, in this paper. The PHYDYAS prototype filter is defined as follows \cite{PHYDYAS,FBMC-book}:
\begin{equation}\label{PHYDYAS}
g(t)=
\begin{cases}
\frac{1}{\sqrt{A}}\left ( 1+2\sum\limits_{k=1}^{K-1}(-1)^k F_k \cos\left(2\pi \frac{kt}{KT} \right ) \right),\\
~~~~~~~~~~~~~~~~~~~~~~~~~~~~~\text{for}~t\in[0,KT],\\
0,~~~~~~~~~~~~~~~~~~~~~~~~~~~\text{otherwise},
\end{cases}
\end{equation}
where $F_0=1,F_1=0.97196,F_2=1/\sqrt{2},F_3=\sqrt{1-F^2_1}$, and $A=KT\left ( 1+2 \sum_{k=1}^{K-1}F^2_k\right )$. The Hermite prototype filter is expressed as \cite{Hermite}
\begin{equation}\label{Hermite}
\begin{split}
  & g(t) \\
= & \begin{cases}
\frac{1}{\sqrt{T}}e^{-2\pi\left ( \frac{t-KT/2}{T}\right )^2}\sum\limits_{k\in \left \{\substack {0,4,8,\\ 12, 16, 20}\right \}}\beta_k H_k\left (2\sqrt{\pi}\frac{t-KT/2}{T} \right ),\\
~~~~~~~~~~~~~~~~~~~~~~~~~~~~~~~~~~\text{for}~t\in[0,KT],\\
0,~~~~~~~~~~~~~~~~~~~~~~~~~~~~~~~~\text{otherwise},
\end{cases}
\end{split}
\end{equation}
where $\{H_k(\cdot)\}$ denote Hermite polynomials and
\begin{equation}
\begin{array}{ll}
\beta_0=1.412682577, & \beta_{4}=-3.0145\times 10^{-3},\\
\beta_{8}=-8.8041\times 10^{-6}, & \beta_{12}=-2.2611\times 10^{-9},\\
\beta_{16}=-4.4570 \times 10^{-15}, & \beta_{20}=1.8633\times 10^{-16}.
\end{array}
\end{equation}
It is noted that the peak of the above PHYDYAS prototype filter is achieved at $t=KT/2=2T$. The same can be said for the Hermite prototype filter defined in (\ref{Hermite}). We will use this setting to study the preamble PAPR in Section III.

\subsection{Introduction to Golay Complementary Pair (GCP)}
Denote by $\rho_{\textbf{c}}(\tau)$ the aperiodic auto-correlation function (AACF) of a complex sequence $\textbf{c}=[c_0,c_1,\ldots,c_{M-1}]$, which is defined as $\rho_{\textbf{c}}(\tau)=\sum_{m=0}^{M-1-m}c_m c^*_{m+\tau}$ for $0\leq \tau\leq M-1$.
%\begin{equation}
%\rho_{\textbf{c}}(\tau)=
%\begin{cases}
%\sum\limits_{m=0}^{M-1-m}c_m c^*_{m+\tau},~~&\text{if}~0\leq \tau\leq M-1;\\
%\rho^*_{\textbf{c}}(-\tau),~~&\text{if}~1-M\leq \tau<0;\\
%0,~~&\text{otherwise}.
%\end{cases}
%\end{equation}
Let $(\textbf{c},\textbf{d})$ be a pair of sequences, both of length $M$. $(\textbf{c},\textbf{d})$ is called a GCP \cite{Golay1961} if $\rho_{\textbf{c}}(\tau)+\rho_{\textbf{d}}(\tau)=0$ for any $\tau\neq0$. Each constituent sequence in a GCP is called a Golay sequence. %Note that compared to conventional one-dimensional sequences, the two constituent sequences in a GCP work in a cooperative way to ensure that their out-of-phase aperiodic autocorrelations sum to zero.

%Let $\phi_{\textbf{C}}(k)=\sum_{k=0}^{N-1}C_kC^*_{n+k~\text{mod}~N}$ be the periodic autocorrelation function (PACF) of $\textbf{C}$ at time-shift $k$. Clearly, $\phi_{\textbf{C}}(k)+\phi_{\textbf{D}}(k)=0$ for any $k\neq0~(\text{mod}~N)$ if $(\textbf{C},\textbf{D})$ is a GCP.

Denote by $\mathbb{Z}_Q$ the set of integers modulo $Q$. For $\mathbf{x}=[x_1,x_2,\ldots,x_{\mu}]\in \mathbb{Z}_2^{\mu}$, a generalized Boolean function (GBF) $f(\mathbf{x})$
is defined as a mapping $f: \{0,1\}^{\mu} \rightarrow \mathbb{Z}_Q$. Each variable $x_i$ ($i\in \{1,2,\ldots,\mu \}$) in $\mathbf{x}$ may be regarded as a GBF (see \textit{Example 1}). Let
$[\kappa_1,\kappa_2,\ldots,\kappa_{\mu}]$ be the binary representation of the
(non-negative) integer $\kappa=\sum_{i=1}^{\mu} \kappa_i 2^{i-1}$, with $\kappa_{\mu}$ denoting the most significant bit and $0\leq \kappa \leq 2^\mu-1$. Given $f(\mathbf{x})$, define $f_\kappa\triangleq f(\kappa_1,\kappa_2,\ldots,\kappa_{\mu})$ and its associated sequence
\begin{displaymath}
\mathbf{{f}}\triangleq\Bigl [f_0, f_1, \ldots, f_{2^\mu-1} \Bigl ].
\end{displaymath}

It is stressed that each realization of $f_{\kappa}$ is obtained by setting $x_1=\kappa_1,x_2=\kappa_2,\ldots,x_{\mu}=\kappa_{\mu}$ into the GBF $f$, for a specific $\kappa$. Hence, if $f=x_t$, where $1\leq t \leq \mu$, the associated sequence $\mathbf{f}$ is the vector formed by $f_\kappa=\kappa_t$, when $\kappa$ ranges from $0$ to $2^\mu-1$. In this case, we denote the corresponding $\mathbf{f}$ by $\mathbf{x}_t$, i.e., $\mathbf{f}=\mathbf{x}_t$. Similarly, if $f=x_{t_1}x_{t_2}$ ($t_1\neq t_2$), we write $\mathbf{f}=\mathbf{x}_{t_1}\mathbf{x}_{t_2}$, a vector obtained by element-wise multiplication of $\mathbf{x}_{t_1}$ and $\mathbf{x}_{t_2}$.

We present the following example to illustrate the GBFs defined above. One can find it useful in understanding the GCP construction in \textit{Lemma \ref{PSK_GDJ constr_4GCP}} below.
\begin{example}
Let $\mu=3$ and $Q=4$. The associated sequences of GBFs $1,x_1,x_3,2x_1x_3+1$ are shown in (\ref{equ4example1}).%the top of next page.
%\begin{displaymath}
%\begin{split}
%\mathbf{1}  & =[1,1,1,1,1,1,1,1],\\
%\mathbf{x}_1 & =[0,1,0,1,0,1,0,1],\\
%\mathbf{x}_3 & =[0,0,0,0,1,1,1,1],\\
%2\mathbf{x}_1\mathbf{x}_3+\mathbf{1} & =[1,1,1,1,1,3,1,3],
%\end{split}
%\end{displaymath}
\begin{figure*}
\begin{equation}\label{equ4example1}
\begin{array}{cccccccccccc}
\left [ \begin{matrix} \kappa_1 \\ \kappa_2 \\ \kappa_3 \end{matrix}\right ] & = & &\left [ \begin{matrix} 0 \\ 0 \\ 0 \end{matrix}\right ] &  \left [ \begin{matrix} 1 \\ 0 \\ 0 \end{matrix}\right ] &  {\color{blue}\left [ \begin{matrix} 0 \\ 1 \\ 0 \end{matrix}\right ]} & \left [ \begin{matrix} 1 \\ 1 \\ 0 \end{matrix}\right ] & {\color{red}\left [ \begin{matrix} 0 \\ 0 \\ 1 \end{matrix}\right ]} &  \left [ \begin{matrix} 1 \\ 0 \\ 1 \end{matrix}\right ] &  \left [ \begin{matrix} 0 \\ 1 \\ 1 \end{matrix}\right ] & \left [ \begin{matrix} 1 \\ 1 \\ 1 \end{matrix}\right ] & \\ \hline \hline
\mathbf{1}  & = &[ &1&1&1&1&1&1&1&1&],\\
\mathbf{x}_1 & =&[&0&1&{\color{blue}0}&1&{\color{red}0}&1&0&1&],\\
\mathbf{x}_3 & =&[&0&0&{\color{blue}0}&0&{\color{red}1}&1&1&1&],\\
2\mathbf{x}_1\mathbf{x}_3+\mathbf{1} & = &[& 1&1&1&1&1&3&1&3&].
\end{array}
\end{equation}
\end{figure*}
%respectively.
When $\kappa=2$, for instance, its binary representation is $[0,1,0]$. Hence, both the third entries of $\mathbf{x}_1$ and $\mathbf{x}_3$ take identical zero. When $\kappa=4$, its binary representation is $[0,0,1]$ and therefore, the fifth entries of $\mathbf{x}_1$ and $\mathbf{x}_3$ are zero and one, respectively. $2\mathbf{x}_1\mathbf{x}_3+\mathbf{1}$ is obtained from two times the element-wise product between $\mathbf{x}_1$ and $\mathbf{x}_3$, followed by addition with $\mathbf{1}$. As observed from (\ref{equ4example1}), the sequence of GBF $x_1$ is in fact given by that of $\kappa_1$, when $\kappa$ ranges from 0 to 7. Similarly, the sequence of GBF $x_3$ is given by that of $\kappa_3$.
%\begin{displaymath}
%\begin{array}{ccl}
%\mathbf{1}   &=&(1,1,1,1,1,1,1,1),\\
%\mathbf{x}_1 &=&(0,1,0,1,0,1,0,1),\\
%\mathbf{x}_3 &=&(0,0,0,0,1,1,1,1),\\
%\mathbf{x}_1\mathbf{x}_3+\textbf{1}&=&(1,1,1,1,1,0,1,0),
%\end{array}
%\end{displaymath}
%respectively.
\end{example}

Next, we present in \textit{Lemma 1} the GCP construction proposed by Davis and Jedwab in \cite{Davis-Jedwab1999}. We will use it to prove in \textit{Remark 3} that a special linear phase transform, pertinent to the FBMC modulation when a binary Golay sequence is adopted as a preamble, of a binary GCP will also be a GCP (quaternary).
%\vspace{0.1in}
\begin{lemma} \label{PSK_GDJ constr_4GCP}(Davis-Jedwab Construction of GCP \cite{Davis-Jedwab1999})
{Let}
\begin{equation}\label{f_4GDJ_GCP}
f(\mathbf{x})\triangleq \frac{Q}{2} \sum \limits_{k=1}^{\mu-1}x_{\pi(k)} x_{\pi(k+1)} + \sum
\limits_{k=1}^{\mu} b_k x_k+b,
\end{equation}
where $\pi$ is a permutation of the set $\{1,2,\ldots,\mu\}$, and $b_k,b\in\mathbb{Z}_{Q}$ ($Q$ even integer).
Then, for any $b'\in \mathbb{Z}_{Q}$, $\mathbf{{f}}~~\text{and}~~\mathbf{{f}}+\frac{Q}{2}{\mathbf{x}}_{\pi(1)}+b'\cdot \mathbf{1}$
%\begin{equation}
%\textit{\textbf{f}}~~\text{and}~~\textit{\textbf{f}}+\frac{q}{2}\textit{\textbf{x}}_{\pi(1)}+c'\cdot \textbf{1}
%\end{equation}
form a GCP over $\mathbb{Z}_{Q}$ of length $2^\mu$. %$\xi=\exp(2\pi\sqrt{-1}/{2^h} )$.
\end{lemma}

\begin{example}
In the context of \textit{Lemma \ref{PSK_GDJ constr_4GCP}}, let $Q=2$, $\mu=4$, and $b=b'=0$. Consider $\pi=[2,3,4,1]$ and $[b_1,b_2,b_3,b_4]=[1,1,0,1]$. Converting from $\mathbb{Z}_2$ to the complex domain, we obtain the length-16 GCP $(\textbf{c},\textbf{d})$ which is shown below, where $``+"$ and $``-"$ stand for $+1$ and $-1$, respectively.
\begin{displaymath}
\left [
\begin{matrix}
\textbf{c}\\
\textbf{d}
\end{matrix}
\right ]=\left[ \begin{matrix}
+--+-+-+++--++++\\
-+-++--+------++
          \end{matrix} \right].
\end{displaymath}
One can verify that $\rho_{\textbf{c}}(\tau)+\rho_{\textbf{d}}(\tau)=0$ for any $\tau\neq0$.
\end{example}

\section{Low-PAPR Preamble Design From Golay Sequences}

%To illustrate the PAPR analysis for preambles,
\subsection{A Generic Framework for Preamble PAPR Analysis}
\begin{figure*}[htbp]
  \centering
  \includegraphics[trim=0.25cm 0.75cm 0.25cm 0.85cm, clip=true, width=6in]{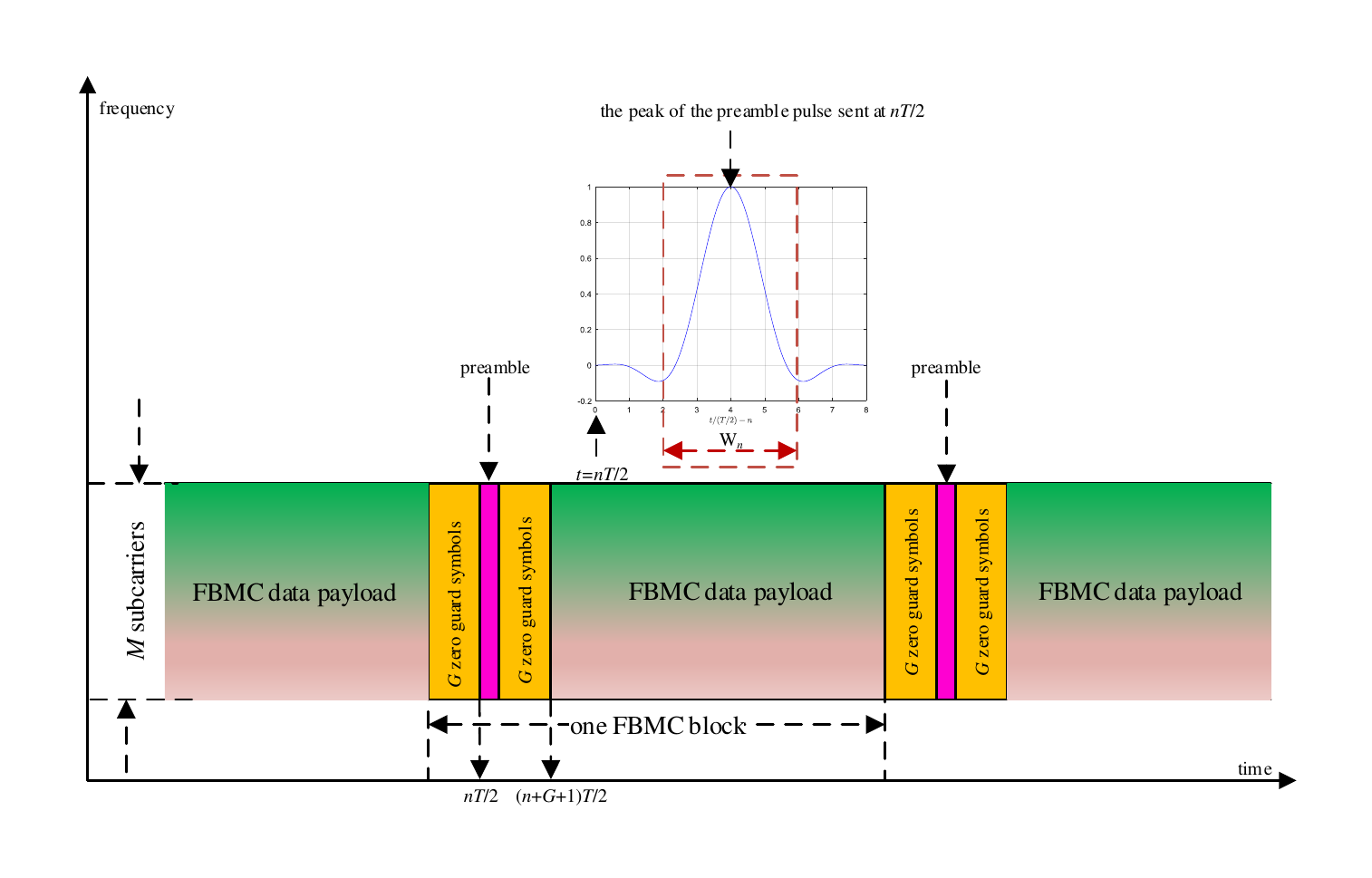}\\
  \caption{FBMC frame structure with zero guard symbols, preambles, and data payloads, where the number of subcarriers is $M$.}
  \label{preamble_structure}
\end{figure*}

Let $\textbf{c}=[c_0,c_1,\ldots,c_{M-1}]$ be a real-valued preamble. This preamble will be transmitted as the $n$-th FBMC symbol, i.e., $a_{m,n}=c_m$ for $0\leq m \leq M-1$, where $a_{m,n}$ is defined in (\ref{TxSIG}). We consider consecutive multi-block transmission of FBMC signals, where a preamble is inserted at the beginning of each block. Moreover, for interference suppression/mitigation, we assume that there are $G$ zero guard symbols placed at the front of this preamble, and another $G$ zero guard symbols at the back, i.e., $a_{m,n-p}=a_{m,n+p}=0$ holds for all $m\in\{0,1,\ldots,M-1\}$ and $p\in \{1,2,\ldots, G\}$. %The first $G$ zero guard symbols are used to prevent the interference from the preceding FBMC block, whereas the second are used to suppress the preamble's interference to the data payload of the current block.
The frame structure considered in this paper is shown in Fig. \ref{preamble_structure}.% below.

Let
\begin{equation}\label{time_win_equ}
\text{W}_n=\left \{t \Bigl |(n+2)T/2\leq t < (n+6)T/2 \right \}
\end{equation}
be a time window centered around $t=(n+4)T/2$ and with duration of $2T$ (i.e., duration of four FBMC symbols). It should be stressed that the time window defined in (\ref{time_win_equ}) is an observation window for preamble PAPR analysis and won't be applied as a prototype filter at the transmitter. As seen in Fig. \ref{preamble_structure}, the peak of the preamble pulse sent at $t=nT/2$ also appears at the $t=(n+4)T/2$. For the PHYDYAS and Hermite prototype filters defined in (\ref{PHYDYAS}) and (\ref{Hermite}), respectively, it can be verified that $99\%$ of the preamble energy is located over the time window $\text{W}_n$. Having this in mind, the preamble PAPR is defined as follows:
\begin{equation}\label{PAPR_def}
\text{PAPR}_{\textbf{c}}\triangleq \max\limits_{t\in \text{W}_n} \frac{|s(t)|^2}{P_{\text{avg}}},
\end{equation}
where $P_{\text{avg}}\triangleq\mathbb{E}[|s(t)|^2]$. For ease of presentation, ${|s(t)|^2}/{P_{\text{avg}}}$ ($t\in \text{W}_n$) is called the instantaneous-to-average power ratio (IAPR) of preamble $\textbf{c}$. Clearly, the preamble PAPR can be obtained by taking the maximum IAPR over the time window $\text{W}_n$. We assume that the OQAM data symbols $a_{m,n'}$ ($n'\neq n$) are independent and identically distributed (real-valued) binary variables with zero mean. Furthermore, we assume that every non-zero FBMC symbol (including the preamble) has energy $M$, i.e., $\sum_{m=0}^{M-1}|a_{m,n'}|^2=M$. The mean power, the denominator of (\ref{PAPR_def}), can be expressed as
\begin{equation}\label{PAPR_def2}
\begin{split}
P_{\text{avg}} & =\frac{2}{T}\int\limits_{0}^{T/2} \mathbb{E}_{a_{m,n'}}[|s(t)|^2] dt\\
                     & = \frac{2}{T} \int \limits_{0}^{T/2} \sum_{n'\in\mathbb{Z}}\left [ \sum\limits_{m=0}^{M-1}|a_{m,n'}|^2\right ]g^2(t-n'T/2) dt\\
                     & = \frac{2M}{T} \int \limits_{0}^{T/2} \sum_{n'\in\mathbb{Z}} g^2(t-n'T/2) dt\\
                     & = \frac{2M}{T},
\end{split}
\end{equation}
where the unit energy property of $g(t)$ is used in the last step of the above derivation. It is noted that the mean power derived in (\ref{PAPR_def2}) is evaluated for the data payload part whose number of symbols is assumed to be sufficiently large. This should not be confused with the mean power over $\text{W}_n$.

To study the peak power over the time window $\text{W}_n$, let us expand $s(t)$ as follows:
\begin{equation}\label{st_expd_equ}
\begin{split}
s(t) & = \sum\limits_{m=0}^{M-1}a_{m,n}g_{m,n}(t)+ \sum\limits_{n'\neq n}\sum\limits_{m=0}^{M-1}a_{m,n'}g_{m,n'}(t)\\
     & = \sum\limits_{m=0}^{M-1}c_{m}g_{m,n}(t)+ \sum\limits_{n'\neq n}\sum\limits_{m=0}^{M-1}a_{m,n'}g_{m,n'}(t).
\end{split}
\end{equation}
For fixed $t$, it is noted that the first summation term in the right-hand side (RHS) of (\ref{st_expd_equ}) is a constant (as $\textbf{c}$ is a preamble), while the second is formed by the summation of a large number of complex-valued random variables $a_{m,n'}g_{m,n'}(t)$. For large $M$, by the central limit theorem, the second term can be well characterized by a complex variable $Z$, where $\Re \left\{Z\right\} \thicksim \mathcal{N}(0,\sigma^2)$ and $\Im\left\{Z\right\}\thicksim \mathcal{N}(0,\sigma^2)$ are statistically independent normal random variables with
\begin{equation}\label{sigma2_equ}
\begin{split}
\sigma^2 & = \frac{1}{2}\cdot\mathbb{E}_{a_{m,n'}}\left \{ \left | \sum\limits_{n'\neq n}\sum\limits_{m=0}^{M-1}a_{m,n'}g_{m,n'}(t) \right |^2\right \}\\
         & = \frac{M}{2} \cdot \sum\limits_{n'\neq n}\text{Ind}(a_{m,n'}\neq 0)g^2(t-n'T/2).
\end{split}
\end{equation}
Here, $\text{Ind}(a_{m,n'}\neq 0)=1$ if $a_{m,n'}\neq 0$ holds and $0$ otherwise. In the latter case, $\text{Ind}(a_{m,n'}\neq 0)=0$ holds for every zero guard symbol, i.e., $n'\in \{n\pm1,n\pm2,\ldots,n\pm G\}$. It is stressed that $\sigma^2$ is a function of $G$ and $t$. In general, a larger $G$ will lead to smaller
$\max\{\sigma^2:t\in\text{W}_n\}$, implying less data interference. It turns out that, for given $t$, $x\triangleq |s(t)|\geq 0$ is characterized by the Rician distribution with the following probability density function (PDF):
\begin{equation}
f(x|\nu,\sigma)=\frac{x}{\sigma^2} \exp\left ( -\frac{x^2+\sigma^2}{2\sigma^2}\right )I_0\left( \frac{x \nu}{\sigma^2}\right),
\end{equation}
where $I_0(\cdot)$ is the modified Bessel function of the first kind with order zero and
\begin{equation}\label{defi_nu_equ}
\nu=\left | \sum\limits_{m=0}^{M-1}c_{m}g_{m,n}(t) \right |.
\end{equation}
Here, $\nu$ gives the magnitude of the instantaneous preamble waveform at time $t$, when the data interference is neglected. As will be shown in (\ref{nu_equ}), $\nu$ is also determined by the prototype filter $g(t)$ and preamble sequence $\mathbf{c}$, yet has nothing to do with $G$. Note that the cumulative density function (CDF) of $x$ is \cite{DigComm-book}
\begin{equation}
1-Q_1\left ( \frac{\nu}{\sigma},\frac{x}{\sigma}\right ),
\end{equation}
where $Q_1(\cdot,\cdot)$ is the standard (first-order) Marcum Q-function:
\begin{equation}
Q_1(a,b)\triangleq \int\limits_{b}^{+\infty}x\exp\left ( -\frac{x^2+a^2}{2}\right )I_0(ax) dx.
\end{equation}

Therefore, the probability that the preamble IAPR is not less than $\alpha$ can be expressed as follows:
\begin{equation}\label{PAPR_equ}
\begin{split}
\text{Pr}\left \{ \frac{|s(t)|^2}{P_{\text{avg}}}\geq \alpha \right \} &= \text{Pr}\left \{ |s(t)|\geq \sqrt{\alpha\cdot P_{\text{avg}}} \right \}\\
                                                                       &= Q_1\left ( \frac{\nu}{\sigma},\frac{\sqrt{\alpha\cdot P_{\text{avg}}}}{\sigma}\right ).
\end{split}
\end{equation}

%Note that each argument of $Q_1$ in the above equation is a function of $t$. Hence, %the probability that $\left \{ \text{PAPR}_{\mathbf{C}}\geq \alpha \right \}$ can be upper bounded by
%\begin{equation}\label{PAPR_equ}
%\begin{split}
%\text{Pr}\left \{ \text{PAPR}_{\mathbf{C}}\geq \alpha \right \} & =1-\text{Pr}\left \{ \frac{|s(t)|^2}{P_{\text{avg}}}< \alpha ~\text{for all}~t\in\text{W}_n\right \}\\
%                                                                & \leq 1-\prod_{t\in \text{W}_n} \left (1-Q_1\left ( \frac{\nu}{\sigma},\frac{\sqrt{\alpha\cdot P_{\text{avg}}}}{\sigma}\right ) \right ),
%\end{split}
%\end{equation}
%where the upper bound is obtained by viewing $\left \{ {|s(t)|^2}/{P_{\text{avg}}}\geq \alpha \right \}$ as independent events over $t\in\text{W}_n$. On the other hand,
%
%$\text{Pr}\left \{ \text{PAPR}_{\mathbf{C}}\geq \alpha \right \}$ should be not less than any $\text{Pr}\left \{ {|s(t)|^2}/{P_{\text{avg}}}\geq \alpha \right \}$.
%By (\ref{PAPR_equ}), we have
%\begin{equation}\label{PAPR_lwrbd_equ}
%\begin{split}
%\text{Pr}\left \{ \text{PAPR}_{\mathbf{C}}\geq \alpha \right \} & = 1-\text{Pr}\left \{ \frac{|s(t)|^2}{P_{\text{avg}}}< \alpha ~\text{for all}~t\in\text{W}_n\right \}\\
%                                                                & \geq 1- \min_{t\in \text{W}_n} \text{Pr}\left \{ \frac{|s(t)|^2}{P_{\text{avg}}}< \alpha \right \}\\
%                                                                & \geq\max_{t\in\text{W}_n} Q_1\left ( \frac{\nu}{\sigma},\frac{\sqrt{\alpha\cdot P_{\text{avg}}}}{\sigma}\right ).
%\end{split}
%\end{equation}

\begin{remark}
It is noted that $Q_1(a,b)$ is strictly increasing with $a$ and strictly decreasing with $b$ \cite{Marcum2010}. Therefore, for fixed $\alpha,G,M,T$, and prototype filter
$g(t)$, (\ref{PAPR_equ}) indicates that it is desirable to have a preamble sequence $\textbf{c}$ with smaller $\nu$ in order to reduce the IAPR. As the PAPR is defined as
the maximum of IAPRs over the time window $\text{W}_n$, a smaller $\nu$ will also lead to a lower PAPR.
\end{remark}

In the sequel, we will use a binary Golay sequence as the FBMC preamble whose corresponding $\nu$ is upper bounded by a certain small value.

\subsection{Application of Golay Sequences as FBMC Preambles}
Recalling $\nu$ defined in (\ref{defi_nu_equ}), we have
\begin{equation}\label{nu_equ}
\begin{split}
\nu & = \left | \sum\limits_{m=0}^{M-1}c_{m}g_{m,n}(t) \right |\\
      & = \left | \sum\limits_{m=0}^{M-1}c_{m}j^{m+n}e^{j2\pi m t/T}g(t-nT/2) \right |\\
      & = g(t-nT/2) \cdot \left | \sum\limits_{m=0}^{M-1}c_{m}j^{m}e^{j2\pi m t/T}\right |
\end{split}
\end{equation}
where $t\in\text{W}_n$.

For any sequence $\textbf{c}$, define $\tilde{\textbf{c}}=[\tilde{c}_0,\tilde{c}_1,\ldots,\tilde{c}_{M-1}]$ where $\tilde{c}_m=c_m j^m$ for $0\leq m \leq M-1$. In fact, $\tilde{\textbf{c}}$ is obtained from a linear phase transform of $\textbf{c}$. To proceed, we need the following remarks on GCPs.

\begin{remark}
 For a given prototype filter and by (\ref{nu_equ}), it is clear that a smaller $\nu$ (and hence lower PAPR by \textit{Remark 1}) is possible provided that $\tilde{\textbf{c}}$ can give rise to a lower instantaneous signal magnitude $\left | \sum_{m=0}^{M-1}c_{m}j^{m}e^{j2\pi m t/T}\right |$ (i.e., lower PAPR of $\tilde{\textbf{c}}$) in conventional OFDM modulation.
\end{remark}

\textit{Remark 2} is the key which motivates us to consider Golay sequences as FBMC preambles.

\begin{remark}\label{rmk_GCP_1}
Suppose $(\textbf{c},\textbf{d})$ is a binary GCP obtained by setting $Q=2$ in \textit{Lemma \ref{PSK_GDJ constr_4GCP}}. Then, $(\tilde{\textbf{c}},\tilde{\textbf{d}})$ will be a quaternary GCP.
\end{remark}
\begin{proof}
%The proof of this remark can be found in \cite{ZL-CPM-2018}. For self-containment, we restate the proof here.
%Next, we will show $\gamma_{\textbf{C}}\triangleq\left\{\gamma_{C,n}\right \}_{n=0}^{N-1}$ and $\gamma_{\textbf{D}}\triangleq\left \{\gamma_{D,n}\right \}_{n=0}^{N-1}$ form a quaternary GCP provided that $(\textbf{C},\textbf{D})$ is a binary GCP generated by the Davis-Jedwab construction (see \textit{Lemma \ref{PSK_GDJ constr_4GCP}}).
In the context of \textit{Lemma \ref{PSK_GDJ constr_4GCP}}, let
\begin{equation}
f=\sum \limits_{k=1}^{\mu-1}x_{\pi(k)} x_{\pi(k+1)} + \sum
\limits_{k=1}^{\nu} b_k x_k+b~~(\text{mod}~2),
\end{equation}
and $f+x_{\pi(1)}+b'~(\text{mod}~2)$ be the GBFs of $\textbf{c}$ and $\textbf{d}$, respectively. Lifting these two GBFs from $\mathbb{Z}_2$ to $\mathbb{Z}_4$ and noting that $\kappa=\sum_{i=1}^{\mu} \kappa_i 2^{i-1}$ ($0\leq \kappa \leq 2^\mu-1$), the $\kappa$-th entries of $\tilde{\textbf{c}}$ and $\tilde{\textbf{d}}$ can be expressed as
\begin{equation}
\begin{split}
\tilde{f}_{\kappa} & =2\sum \limits_{i=1}^{\mu-1}\kappa_{\pi(i)} \kappa_{\pi(i+1)} + 2\sum
\limits_{i=1}^{\mu} b_i \kappa_i\\
& ~~~~+2b+\sum_{i=1}^{\mu} \kappa_i 2^{i-1}~~(\text{mod}~4)\\
   & =2\sum \limits_{i=1}^{\mu-1}\kappa_{\pi(i)} \kappa_{\pi(i+1)} + 2\sum
\limits_{i=1}^{\nu} b_i \kappa_i\\
   & ~~~~+2b+2\kappa_2+\kappa_1~~(\text{mod}~4)\\
   & =2\sum \limits_{i=1}^{\mu-1}\kappa_{\pi(i)} \kappa_{\pi(i+1)} + \sum
\limits_{i=3}^{\mu} (2b_i) \kappa_i\\
   & ~~~~+(2+2b_2)\kappa_2+(1+2b_1)\kappa_1+2b~~(\text{mod}~4),
\end{split}
\end{equation}
and $\tilde{f}_{\kappa}+2\kappa_{\pi(1)}+2b'~(\text{mod}~4)$, respectively. It is clear that $\tilde{f}_{\kappa},\tilde{f}_{\kappa}+2\kappa_{\pi(1)}+2b'~(\text{mod}~4)$ satisfy the GBF forms in \textit{Lemma \ref{PSK_GDJ constr_4GCP}} with $Q=4$. Thus, $\tilde{\textbf{c}}$ and $\tilde{\textbf{d}}$ form a quaternary GCP.
\end{proof}
\begin{remark}\label{rmk_GCP_2}
For any polyphase Golay sequence $\textbf{c}$ with energy of $M$, its instantaneous signal magnitude under conventional OFDM modulation is upper bounded by $\sqrt{2M}$, i.e.,
\begin{equation}\label{GolaySeq_PAPR}
\left | \sum\limits_{m=0}^{M-1}c_{m}e^{j2\pi m t/T}\right | \leq \sqrt{2M}.
\end{equation}
\end{remark}
\begin{proof}
See \cite{Davis-Jedwab1999}.
\end{proof}

As a matter of fact, numerical experiments show that the upper bound of (\ref{GolaySeq_PAPR}) can be well approached \cite{Davis-Jedwab1999,Wang2014}. %In particular, for binary Golay sequence with $\log_2(M)$ being an odd integer, the upper bound of (\ref{GolaySeq_PAPR}) can be met with equality \cite{Wang2014}.

From now on, let us assume that the preamble $\textbf{c}$ is a binary Golay sequence. By \textit{Remark \ref{rmk_GCP_1}}, $\tilde{\textbf{c}}$ is also a Golay sequence. Substituting $\tilde{\textbf{c}}$ into \textit{Remark \ref{rmk_GCP_2}}, it follows from (\ref{nu_equ}) that% $\nu$ can be upper bounded as follows.
\begin{equation}\label{nu_equ2}
\nu \leq \sqrt{2M} \cdot g(t-nT/2).
\end{equation}

As shown in (\ref{PAPR_equ}), the IAPR is also influenced by $\sigma^2$, the variance of the data interference [see (\ref{sigma2_equ})]. The amount of $\sigma^2$ can be controlled by $G$, the number of zero guard symbols. When $G=1$, as will be demonstrated later, the preamble optimization will not be effective as the preamble PAPR will be strongly affected by data interference. On the other hand, as $G$ grows to a large number, e.g., $G=3$, the preamble PAPR can be well upper bounded by a small value. In particular, when $\sigma=0$, we have $s(t)=\sum_{m=0}^{M-1}c_{m}g_{m,n}(t)$. Hence, we can adopt the following simplified analysis to evaluate the preamble PAPR:
\begin{equation}\label{PAPR_zero_sigma}
\text{PAPR}^{\{\sigma=0\}}_{\textbf{c}}= \max\limits_{t\in \text{W}_n} \frac{\nu^2}{P_{\text{avg}}}\leq T \max_{t\in \text{W}_n}g^2(t-nT/2).
\end{equation}

\begin{remark}
When a Golay sequence is adopted as an FBMC preamble, (\ref{PAPR_zero_sigma}) indicates that a prototype filter with better time-localization gives rise to higher PAPR. Thus, by the Heisenberg-Gabor principle \cite{Boashash2003}, a prototype filter with better frequency-localization is preferred for lower PAPR.
\end{remark}

We will illustrate the observation in \textit{Remark 5} through the PAPR analysis for the PHYDYAS and Hermite prototype filters below.

For the PHYDYAS prototype filter defined in (\ref{PHYDYAS}), recall that the peak value of $g(t)$, where $t\in[0,KT]$, is achieved at $t=0$. By (\ref{PAPR_zero_sigma}), we have
\begin{equation}\label{PAPR_zero_sigma2}
\text{PAPR}^{\{\sigma=0\}}_{\textbf{c},\text{PHYDYAS}}\leq \frac{\left (1+ 2\sum\limits_{k=1}^{K-1}F_k \right)^2}{K\left ( 1+2 \sum\limits_{k=1}^{K-1}F^2_k\right )}\thickapprox 1.635 ~\text{dB}\footnote{For the PHYDYAS prototype filter with overlapping factor of $K=3$ \cite{FBMC-book}, $F_0=1,F_1=0.911438,F_2=0.411438$, it can be readily shown that this upper bound is 1.693 dB, slightly higher than that of $K=4$ owing to its worse frequency-localization.}.% 1.457.
\end{equation}

For the Hermite prototype filter defined in (\ref{Hermite}), similarly, we have
\begin{equation}\label{PAPR_zero_sigma3}
\text{PAPR}^{\{\sigma=0\}}_{\textbf{c},\text{Hermite}}\leq \left (\sum_{k\in \left \{\substack {0,4,8,\\ 12, 16, 20}\right \}}\beta_k H_k(0) \right )^2\thickapprox 2.673 ~\text{dB}.%1.851.
\end{equation}

\begin{figure}%[htbp]
  \centering
  \includegraphics[trim=2.5cm 8.5cm 2.5cm 8.5cm, clip=true, width=3in]{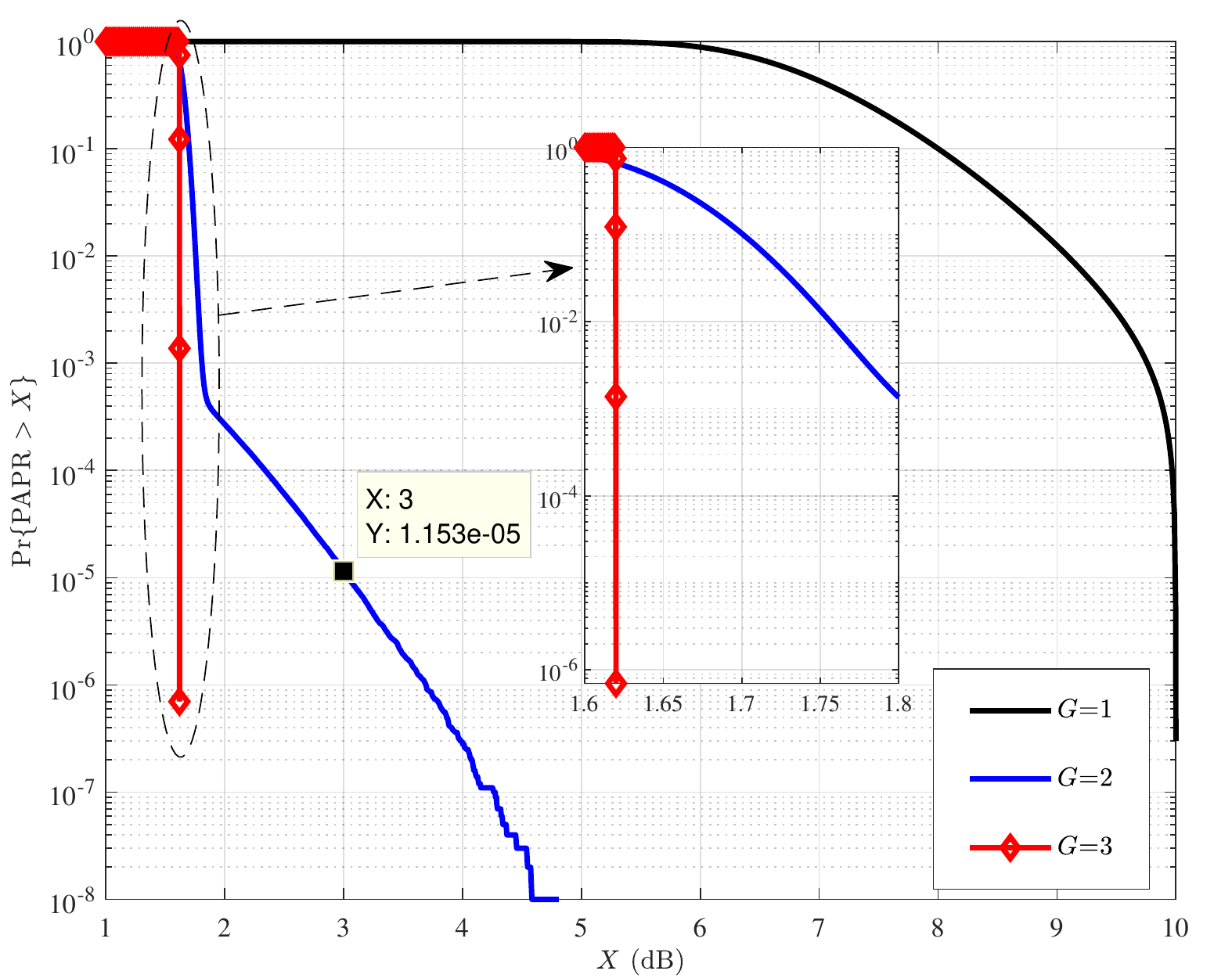}\\
  \caption{Preamble PAPR CCDFs for the PHYDYAS prototype filter with $G\in \{1,2,3\}$.}
  \label{PAPR_PHYDYAS}
\end{figure}

\begin{figure}%[htbp]
  \centering
  \includegraphics[trim=2.5cm 8.5cm 2.5cm 8.5cm, clip=true, width=3in]{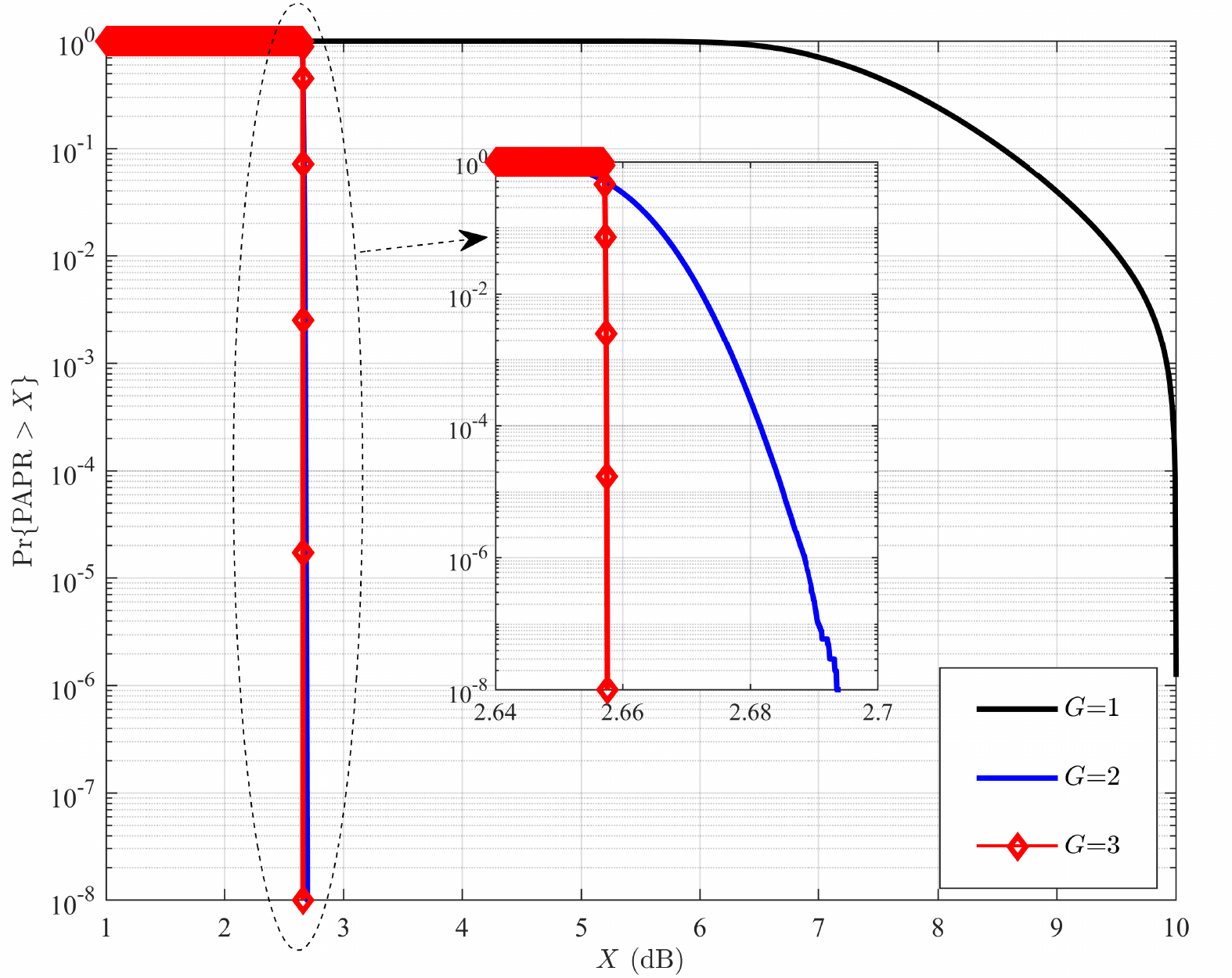}\\
  \caption{Preamble PAPR CCDFs for the Hermite prototype filter with $G\in \{1,2,3\}$.}
  \label{PAPR_Hermite}
\end{figure}

In Figs. \ref{PAPR_PHYDYAS} and \ref{PAPR_Hermite}, we show the complementary cumulative density functions (CCDFs) of preamble PAPRs, i.e., $\text{Pr}\{\text{PAPR}>X\}$, for both the PHYDYAS and Hermite prototype filters with $G\in \{1,2,3\}$, estimated through simulations. We can make the following observations:
\begin{enumerate}
\item  For $G=1$, as expected, the preamble PAPR for each prototype filter can take value much larger than 3 dB with high probability.
\item  For $G=2$, the preamble PAPR of the PHYDYAS prototype filter is lower than 3 dB with probability of
\begin{displaymath}
1-1.153\times10^{-5}>99.99\%.
\end{displaymath}
In contrast, the preamble PAPR of the Hermite prototype filter is upper bounded by 2.7 dB.
\item For $G=3$, one can see that the preamble PAPRs for the two prototype filters are close to (and upper bounded by) 1.635 dB [see (\ref{PAPR_zero_sigma2})] and 2.673 dB\footnote{Fig. 2 shows that the preamble PAPR of the Hermite prototype filter for $G=3$ is upper bounded by 2.660 dB.}  [see (\ref{PAPR_zero_sigma3})], respectively. %A $G$ value larger than 3 may not bring in significant
\item The preamble PAPR CCDFs of the Hermite prototype filter for $G=2$ and $G=3$ are much closer than those of the PHYDYAS prototype filter, owing to the
better time-localization of the former \cite{NSR2017}.
\end{enumerate}

Next, we show that \textit{sparse} preambles \cite{KKRT2010} with low PAPRs can be synthesized from binary Golay sequences. For a ``seed" GCP $(\textbf{c},\textbf{d})$ of length $2^\mu$, it can be easily proved that \cite{Fan-book}
\begin{displaymath}
\left (\textbf{c}\otimes \left[1,0^{1\times d}\right],\textbf{d}\otimes \left[1,0^{1\times d} \right] \right )
\end{displaymath}
form a sparse GCP of length $(d+1)2^{\mu}$, where $\otimes$ and $0^{1\times d}$ denote the Kronecker product and a length-$d$ zero row vector, respectively. For an FBMC system with $M$ subcarriers and channel memory length of $L_h$, it is found in \cite{KKRT2010} that the \textit{optimal}\footnote{The \textit{optimality} of this class of sparse preambles is that it is capable of achieving the minimum mean squared error with a given preamble energy.}  sparse preambles constitute $L_h$ isolated pilots which should be equi-spaced and equi-powered. For example, if $M=8,L_h=4$, $[c_0,0,c_1,0,c_2,0,c_3,0]$ will be an optimal sparse preamble provided that $c_0,c_1,c_2,c_3$ take identical magnitude. Although it is known that a lower PAPR is expected for a preamble with larger sparsity, PAPR reduction is not considered as a constraint in the preamble optimization of \cite{KKRT2010}.  In fact, by choosing a seed GCP with $2^\mu\geq L_h$, one can assert that
\begin{equation}\label{GolaySeq_sparsify}
\sqrt{{M}/{2^\mu}}\left (\textbf{c}\otimes \left [1,0^{1\times d} \right ]\right )%~\text{or}~\sqrt{{M}/{2^\mu}}\left (\textbf{d}\otimes \left [1,0^{1\times d} \right ]\right )
\end{equation}
 is an \textit{optimal} sparse preamble but with low PAPR property, where the factor $\sqrt{{M}/{2^\mu}}$ is used to normalize the preamble energy to $M$.

  \begin{figure}%[htbp]
  \centering
  \includegraphics[trim=2cm 8cm 2cm 8cm, clip=true, width=3in]{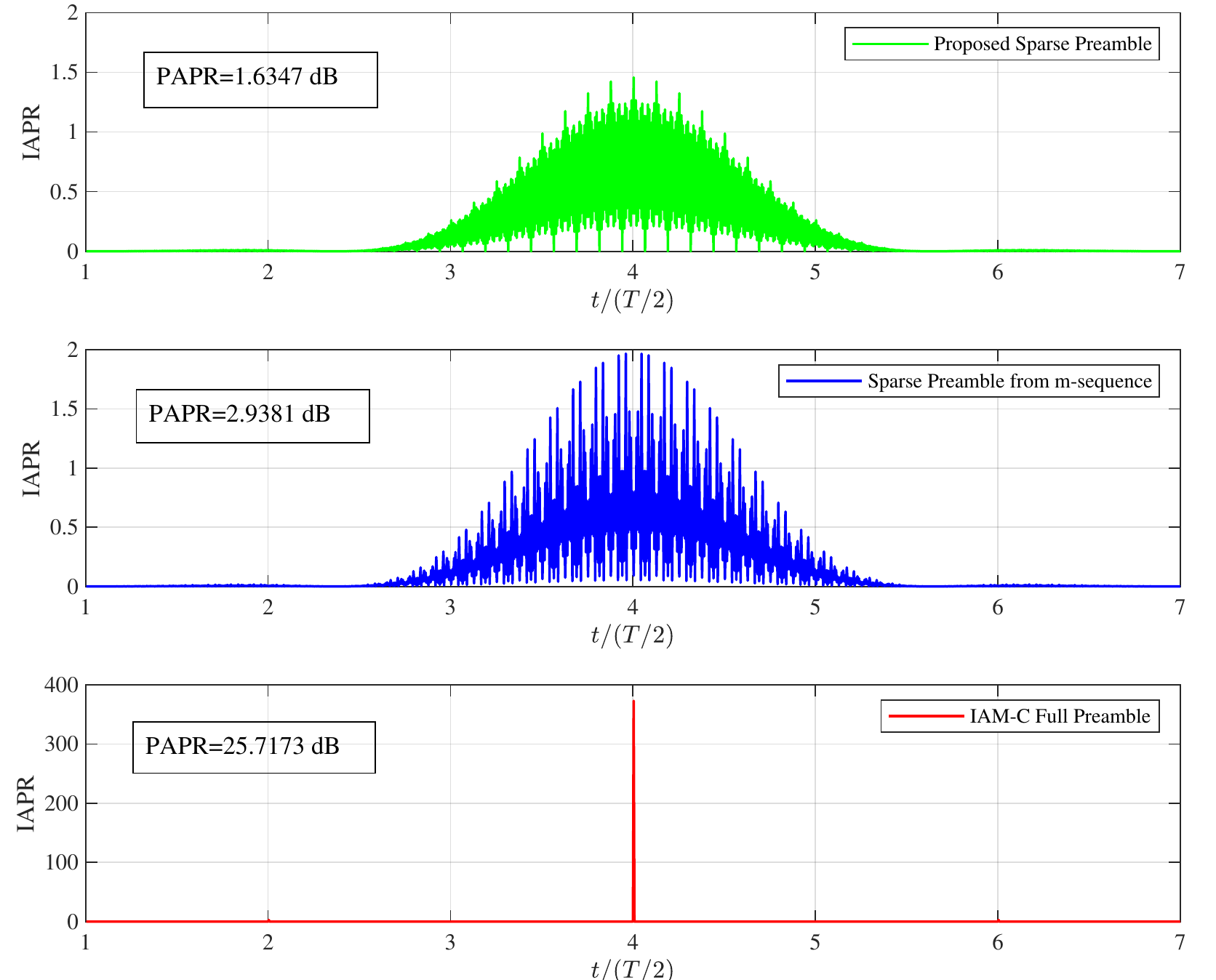}\\
  \caption{PAPR comparison of different preambles using the PHYDYAS prototype filter with $K=4$.}
  \label{PAPRComp}
\end{figure}

In Fig. \ref{PAPRComp}, we compare the PAPRs of different preambles using the PHYDYAS prototype filter defined in (\ref{PHYDYAS}). It is assumed that $G$ is sufficiently large, the number of subcarriers $M$ is equal to $512$ and $L_h=32$. The proposed sparse preamble is built from the length-32 Golay sequence\footnote{One can easily obtain this length-32 Golay sequence by concatenating the two length-16 Golay sequences in \textit{Example 2}. Interested readers may refer to \cite{Golay1961} and \cite{Fan-book} for constructions of long GCPs with various sequence operations.} $[+--+-+-+++--++++-+-++--+------++]$ using (\ref{GolaySeq_sparsify}). The other sparse preamble is generated in a similar manner, but using the length-31 m-sequence $[+----+--+-++--+++++---++-+++-+-]$, which has periodic autocorrelations of $-1$ for all the non-zero time-shifts, concatenated with a $+$. IAM-C is a well-known \textit{optimal} preamble whose entries all take non-zero values\footnote{In \cite{KKRT2010}, such preambles are called ``full".} \cite{Du2008}. It is seen that the PAPR of the proposed preamble is 1.6347 dB\footnote{This is consistent with the PAPR upper bound of 1.635 dB in (\ref{PAPR_zero_sigma2}).}, compared to 2.9381 dB for that of the sparse preamble based on the m-sequence. This shows the strength of Golay sequences over pseudo-random sequences for preamble PAPR reduction in FBMC systems. In sharp contrast, the IAM-C dense preamble has PAPR of 25.7173 dB, which may be unbearable in practical FBMC system design \cite{Levanen2016}.

\section{Conclusions}

In this paper, we have studied the preamble PAPR in FBMC systems employing the PHYDYAS and the Hermite prototype filters, using, as an illustration of the most widely used case, an overlapping factor of 4. We have shown (in \textit{Remark 2}) that an FBMC preamble with low PAPR is possible provided that the linear phase transform (i.e., modulated with $j^m$, where $m$ denotes the subcarrier index) of such a preamble can give rise to a low PAPR in conventional OFDM systems. When the number of zero guard symbols is three or more, it is found that preambles constructed from binary Golay sequences have PAPRs upper bounded by 1.635 dB and 2.673 dB, for the PHYDYAS and the Hermite prototype filters, respectively. With a Golay sequence adopted as an FBMC preamble, we have observed (in \textit{Remark 5}) that a prototype filter with better frequency localization gives rise to lower PAPR. This is evident from the PAPR comparison of the two prototype filters as shown in Figs. 1-2. \textit{Optimal} sparse FBMC preambles have been constructed from binary Golay sequences [see (\ref{GolaySeq_sparsify})] for a practical channel estimation scheme. In the end, we point out that, by a similar analysis, Golay sequences also lead to low-PAPR preambles in FBMC/QAM systems \cite{Farhang2011}.

\section{Acknowledgement}

The authors are deeply indebted to Prof. Eleftherios Kofidis at University of Piraeus for many of his insightful comments.

\end{document}